\documentclass[a4paper,prb,aps,epsfig,twocolumn,showpacs,preprintnumbers,superscriptaddress,float,amsmath,amssymb]{revtex4}
\usepackage{graphicx}
\usepackage{dcolumn}
\usepackage{bm}

\begin{document}

\title{Gate-tunable Kondo resistivity and dephasing rate in graphene studied by numerical renormalization group calculations}

\author{Po-Wei Lo}
\affiliation{Department of Physics and Center for Theoretical Sciences, National Taiwan University, Taipei 10617, Taiwan}
\author{Guang-Yu Guo}
\affiliation{Department of Physics and Center for Theoretical Sciences, National Taiwan University, Taipei 10617, Taiwan}
\affiliation{Graduate Institute of Applied Physics, National Chengchi University, Taipei 11605, Taiwan}
\author{Frithjof B. Anders}
\affiliation{Lehrstuhl f$\ddot ur$ Theoretische Physik II, Technische Universit$\ddot a$t at Dortmund, DE-44221 Dortmund, Germany}

\date{\today}

\begin{abstract}
Motivated by the recent observation of the Kondo effect in graphene in transport experiments, we investigate 
the resistivity and dephasing rate in the Kondo regime due to magnetic impurities in graphene
with different chemial potentials ($\mu$). The Kondo effect due to either carbon vacancies or magnetic adatoms 
in graphene is described by the single-orbital pseudo-gap asymmetric Anderson impurity model 
which is solved by the accurate numerical renormalization group method. 
We find that although the Anderson impurity model considered here is a mixed valence system,
it can be drived into either the Kondo [$\mu > \mu_c$ (critical value) $>0$] or 
mixed valency ($\mu \approx \mu_c$) or empty orbital ($\mu < \mu_c$) regime by a gate voltage,
giving rise to characteristic features in resistivity and dephasing rate in each regime.
Specifically, in the case of $\mu < \mu_c$, the shapes of the resistivity (dephasing rate) 
curves for different $\mu$ are nearly identical. However, as temperature decreases, 
they start to increase to their maxima at 
a lower $T/T_K$ but more rapidly [as $(T_K/T)^{3/2}$] than in normal metals [here $T$ ($T_K$) denotes the (Kondo) temperature]. 
As $T$ further decreases, after reaching the maximum, the dephasing rate drops more quickly than in normal metals,
behaving as $(T/T_K)^3$ instead of $(T/T_K)^2$. Furthermore, the resistivity has a distinct 
peak above the saturation value near $T_K$.
In the case of $\mu > \mu_c$, in contrast, the resistivity curve has an additional broad shoulder above 10$T_K$
and the dephasing rate exhibits an interesting shoulder-peak shape.  
In the narrow boundary region ($\mu \approx \mu_c$), both the resistivity and dephasing rate curves
are similar to the corresponding ones in normal metals. 
This explains the conventional Kondo like resistivity from recent experiments on graphene with defects, 
although the distinct features in the resistivity in the other cases ($\mu < \mu_c$ or $\mu > \mu_c$) 
were not seen in the experiments. 
The interesting features in the resistivity and dephasing rate are analysized 
in terms of the calculated $T$-dependent spectral function, correlation self-energy and renormalized impurity level.
 
\end{abstract}

\pacs{72.10.Fk, 72.15.Qm, 72.80.Vp, 73.22.Pr}

\maketitle

\section{Introduction}
Interactions between particles in many-body systems are responsible for the formation of correlated many-body states 
which lead to a plethora of quantum phenomena. Therefore, many-body correlation in quantum systems has been 
one of the most significant topics in condensed matter physics. Graphene, a newly discovered two-dimensional material 
with carbon atoms arranged in a single-layer honeycomb lattice possesing an unusual 
band structure with a linear spectrum, hosts two-dimensional Dirac fermions \cite{Nov04,Nov05,Net09}. It provides 
a great opportunity to investigate novel many-body correlation in low-dimensional Dirac fermion systems.
Furthermore, the easy manipulation of the Fermi level by an applied gate voltage also makes graphene an excellent candidate 
for important technological applications. 

In particular, there has been an increasing interest in low-temperature behaviors of dilute magnetic impurities in 
graphene in recent years\cite{Fri13}. Local magnetic moments in graphene can be created by placing magnetic atoms onto 
the graphene sheet or by introducing point defects in the graphene sheet\cite{Yaz07,Cha08,Weh10,Weh11,Sof12,Nai12}. The 
theoretical model for describing the interactions of local magnetic moments with conduction electrons in metals 
was first formulated by Anderson, known as the Anderson impurity model\cite{And61}. The impurity and conduction electrons
could form an entangled many-body state below the Kondo temperature ($T_K$) in the presence of the hybridizations between them. 
The magnetic moment of the impurity would then be screened by the formation of the many-body singlet, 
known as the Kondo effect\cite{Hew93}.  However, in systems with a pseudogap like graphene, the Kondo effect could be 
significantly different.\cite{Wit90,Che95,Ing96,Bul97,Gon98,Fri04,Che13}. For example, with vanishing density of state 
at the Fermi level, the tendency toward the screened impurity in graphene would get reduced. 
Therefore, the Kondo effect would only take place when the strength of the coupling of the impurity state 
to the conduction electrons ($\Gamma_0$) exceeds a critical value ($\Gamma_c$), resulting in a unusual phase diagram for the system. 
Up to now, the thermodynamical properties of graphene Kondo systems have been extensively studied\cite{Fri13,Gon98,Kan12}. 

However, the thermodynamical properties in graphene with dilute impurities are difficult to probe. On the other hand,
transport properties are easier to measure. Indeed resistivity evidence for the Kondo effect
in graphene with vacancies has been recently reported\cite{Che11}.
By the manipulation of the Fermi level by an applied gate voltage, the carrier density becomes controllable. This provides
a rare opportunity to study the carrier density-tuning of the Kondo effect, which could lead to many interesting 
transport phenomena\cite{Che11,Sen08,Zhu09,Voj10,Com09,Jac10}. 
Under an applied gate voltage, for example, the Fermi level would move away from the Dirac point, and the density of states 
at the Fermi level becomes finite. Therefore, the magnetic moment of the impurity can be screened in the 
low-temperature limit. Although the thermodynamical properties of the system might become similar to that of normal metals,
the impurity spectral function could have unusual characteristics\cite{Zhu09,Voj10}. 
These unusual characteristics would make the behaviors of the transport properties such as resistivity and dephasing rate
become different from that of the conventional Kondo effect.
 
So far, however, there have been few studies on the transport properties in graphene with the Kondo effect\cite{Zhu09, Voj10}
and all these studies are on the tunneling spectra for scanning tunneling spectroscopy. 
In Ref. \onlinecite{Zhu09}, the Anderson impurity model with the on-site Coulomb interaction $U \rightarrow \infty$,
was solved within the slave-boson mean-field approximation, whereas in Ref. \onlinecite{Voj10}, 
the standard spin-1/2 Kondo model was treated using the non-perturbative numerical renormalization group (NRG) method\cite{Wil75}. 
In this work, we systematically study the behaviors of resistivity and dephasing rate in the Kondo regime in
the asymmetric Anderson impurity model with the realistic model parameters for graphene with carbon vacancies or
magnetic adatoms. We use the accurate NRG method.\cite{Wil75}
We find that although the asymmetric Anderson impurity model considered here is a mixed valence system,
it can be drived into either the Kondo or mixed valency or empty orbital regime by a gate voltage,
giving rise to characteristic features in resistivity and dephasing rate in each regime.
The results not only explain the Kondo resistivity observed in recent experiments\cite{Che11} but also 
suggest directions for further transport experiments to identify the exotic features of the Kondo effect in graphene.
This rest of this paper begins with an introduction 
to the structure of graphene with local magnetic moments in Sec. II. This will be followed by a brief description of 
the Anderson impurity model for graphene with carbon vacancies and magnetic adatoms, and the NRG method in Sec. II. 
Then, the calculated resistivity and dephasing rate are presented 
in Sec. III. Recent experiments on the Kondo effect in graphene are commented and further experiments on transport 
properties are suggested in Sec. IV.  Finally, the conclusions that could be drawn from this work are given in Sec. V.

\section{Theory and Computational Method}

\subsection{The Anderson impurity model}
The positions of impurities play an important role in the Kondo physics in graphene. 
Possible absorption sites on graphene sheet can be classified as the top, hollow, and bridge ones. 
The systems with adatoms absorbed on the hollow and bridge sites are the candidates to realize the 
multi-channel Kondo problem\cite{Zhu10}. 
Other possible structures include carbon vacancies and substitutional impurities\cite{Kan12,Uch11}. 
Nevertheless, in the systems with magnetic adatoms absorbed on the hollow and bridge sites and also 
substitutional impurities, the scattering rate is proportional to $|\omega+\mu|^3$ where $\omega$ 
and $\mu$ are the energy and chemical potential, respectively. This suggests that 
the Kondo effect would hardly occur in these systems\cite{Zhu10,Uch11}. 
In this paper, therefore, we consider only the cases of 
carbon vacancies and also magnetic adatoms absorbed on the top-sites which are the single-channel Kondo problems. 

The Anderson impurity model can be wrriten as\cite{Kan12,Zhu10}:
\begin{equation}
\begin{split}
H = \sum_\sigma\epsilon_df_\sigma^\dagger f_\sigma + Uf_\uparrow^\dagger f_\uparrow f_\downarrow^\dagger f_\downarrow\ \ \ \ \ \ \ \ \ \ \ \ \ \ \ \ \ \ \ \ \ \ \ \ \ \\
 + \sum_\sigma\int_{-D}^{D}d\omega |\omega+\mu|g_\sigma(\omega)c_\sigma^{\dagger}(\omega)c_\sigma(\omega)\ \ \ \ \ \ \\
 + \sum_\sigma\int_{-D}^{D}d\omega \sqrt{\frac{\Gamma(\omega)}{\pi D}}[f_\sigma^\dagger c_\sigma(\omega)+c_\sigma^{\dagger}(\omega)f_\sigma],
\end{split}
\end{equation}
where $\sigma = \uparrow, \downarrow$; $\epsilon_d$ is the energy of the impuriy level, and $U$ is the Coulomb 
interaction between the electrons; $f_\uparrow^\dagger$ ($f_\downarrow^\dagger$) and $f_\uparrow$ ($f_\downarrow$) 
are creation and annihilation operators for an electron in the $\uparrow$ ($\downarrow$) impurity state; 
$c_\uparrow^\dagger(\omega)$ [$c_\downarrow^\dagger(\omega)$] and $c_\uparrow(\omega)$ [$c_\downarrow(\omega)$] 
are creation and annihilation operators for an electron in the $\uparrow$ ($\downarrow$) 
conduction states with energy equal to $\omega$; $g_\sigma(\omega)$ is the part of the density of states 
that couples to the impurity state. $D$ is the total band width.
$\Gamma(\omega)$ is the scattering rate and related to $g_\sigma(\omega)$ by 
equation $\Gamma(\omega)=\pi g_\sigma(\omega) |V_{hyb}(\omega)|^2$, 
where $V_{hyb}(\omega)$ is the effective hybridization strength.

In the case of carbon vacancies, the scattering rate $\Gamma(\omega)$ can be written as\cite{Kan12}:
\begin{equation}
\Gamma^{(vac)}(\omega) = \frac{\Omega_0 V_{vac}^2|\omega+\mu|}{2 \hbar^2 v_F^2}[2-J_0(\frac{2}{3}\frac{|\omega+\mu|}{t})],
\end {equation}
where $\Omega_0$, $v_F$, $t$, and $V_{vac}$ are the unit cell area, the Fermi velocity, the hopping energy, 
and the hybridization strength, respectively; $J_0$ is the zeroth Bessel function. Since we only consider 
the cases of small chemical potentials, the value of $(2/3)(|\omega+\mu|/t)$ is small for small $\omega$. 
Therefore, we can expand $J_0$ at $\omega = 0$, and $\Gamma^{(vac)}(\omega)$ can be approximated as:
\begin{equation}
\Gamma^{(vac)}(\omega) = \frac{\Omega_0 V_{vac}^2|\omega+\mu|}{2 \hbar^2 v_F^2}[1+\frac{4}{27}(\frac{|\omega+\mu|}{t})^2].
\end {equation}
Hence the effective hybridization strength can be written as:
\begin{equation}
V_{hyb}^{(vac)}(\omega) = V_{vac}\sqrt{1+\frac{4}{27}(\frac{|\omega+\mu|}{t})^2}.
\end{equation}

In the case of adatoms on the top sites, the tight-binding formalism is used. We only need to consider 
the hybridizations of the impurity with the conduction electrons below the impurity and the next 
nearest neighbors. Thus, the scatterring rate can be written as\cite{Zhu10}:
\begin{equation}
\Gamma^{(ada)}(\omega) = \frac{\Omega_0 V_A^2}{2 \hbar^2 v_F^2}[1-\frac{(\omega+\mu)}{t}\frac{V_B}{V_A}]^2,
\end {equation}
where $V_A$ is the hybridization of the impurity with conduction elecron below the impurity, 
and $V_B$ is the hybridization of the impurity with conduction electrons on the next nearest neighbors. 
In general, $V_B\ll V_A$. The effective hybridization strength can be approximated as:
\begin{equation}
V_{hyb}^{(ada)}(\omega) = V_A [1-\frac{(\omega+\mu)}{t}\frac{V_B}{V_A}].
\end{equation}

In this work, therefore, the following scattering rate is assumed:
$$ \Gamma (\omega)=\left\{
\begin{array}{rcl}
\Gamma_0 |\tilde{\omega}+\tilde{\mu}| s(\tilde{\omega}+\tilde{\mu})     &   & {|\tilde{\omega}+\tilde{\mu}| \leq 1}\\
\Gamma_0 s(\frac{\tilde{\omega}+\tilde{\mu}}{|\tilde{\omega}+\tilde{\mu}|})   &   & {1<|\tilde{\omega}+\tilde{\mu}|,|\tilde{\omega}| \leq \frac{D} {D_{eff}}} \\
0    &    & {|\tilde{\omega}| > \frac{D}{D_{eff}}}
\end{array} \right. $$
where $\Gamma_0 =  \Omega_0 V^2 D_{eff}/2 \hbar^2 v_F^2$; $V$ = $V_{vac}$ ($V_A$) for the case of 
carbon vacancies (adatoms); $\tilde{\omega} = \omega/D_{eff}$ and $\tilde{\mu} = \mu/D_{eff}$;
$D_{eff}$ is the effective band width within which the density of states is approximately proportional 
to $|\omega+\mu|$; $s(x)$ is defined as:\\
$ s(x)=\left\{
\begin{array}{rcl}
1+\alpha x^2     &   & \textrm{for\ carbon\ vacancies}\\
(1-\beta x)^2       &   & \textrm{for\ adatoms}\\
\end{array} \right.$\\
where $\alpha = (4/27)(D_{eff}/t)^2$ and $\beta = (D_{eff}/t)(V_B/V_A)$. 

In this work, we exploit the powerful NRG method\cite{Wil75} 
to solve the Anderson impurity model. In all the present calculations, we use the discretization 
parameter $\Lambda$ = 1.8 and keep 1200 states per NRG iteration so that the obtained
resistivity and dephasing rate converge within 0.1 \%. The valence band-width of the 
graphene band structure is about 20 eV and the linear dispersion extends up to about 2 eV above 
and below the Dirac point\cite{Cha08}. Therefore, we set $D$ = 20 eV and $D_{eff}$ = 2 eV. 
Parameters $\epsilon_d$, $U$ and $\Gamma_0$ ($V$) would depend on the type of
the impurities. Typically, $U$ varies from about 1 to 10 eV, $\Gamma_0$ of the order of 1 eV,
and $\epsilon_d$ is around -1 eV.\cite{Yaz07,Cha08,Weh10,Weh11,Voj10,Kan12} 
We have explored a wide parameter range. The calculated 
thermal properties such as susceptibility, entropy, Kondo temperature  and specific heat, 
are consistent with previous theoretical reports\cite{Gon98,Zhu09,Voj10,Kan12}. 
In this paper, we use parameters $U$ = 10 eV, $\epsilon_d$ = -0.7 eV and $\Gamma_0$ = 1 eV 
(i.e., $V = 2.6$ eV). That is, we consider a mixed valence system 
since $|\epsilon_d|/\Gamma_0 \le 1$\cite{Kri80,Cos94}. 
Interestingly, as will be demonstrated below, this mixed valence graphene system 
can be drived into either the Kondo or mixed valency or empty orbital regime by a gate voltage,
giving rise to contrasting behaviors in resistivity and dephasing rate.

\subsection{Resistivity calculation}
To obtain resistivity and dephasing rate, we first calculate the temperature ($T$) dependent 
single-particle Green's function $G_\sigma^f(\omega,T)$ of the impurity by the NRG method\cite{Wil75}. 
The impurity spectral function is defined 
as $A_\sigma^f (\omega ,T) = -(1/\pi) \textrm{Im}G_\sigma^f (\omega ,T)$, and can be 
calculated directly by the Lehmann representation:
\begin{equation}
\begin{split}
A_\sigma^f (\omega ,T) = \frac{1}{Z(T) }\sum_{r,r'}|M_{r,r'}|^2 (e^{E_r /k_B T}+e^{E_r' /k_B T}) \\
\times\delta (\omega -(E_r' -E_r)).
\end {split}
\end{equation}
where $Z(T)$ is the partition function and $M_{r,r'} = \langle r|f_\sigma |r'\rangle$ is the relevant many-body matrix
element; $|r\rangle$ ($|r'\rangle$) is the many-body eigenstate and $E_r$ ($E_{r}'$) is the corresponding eigenenergy.
The real part of $G_\sigma^f(\omega,T)$ can be obtained via Kramers-Kronig relation. In the present calculations,
the method for improving the resolution of $A_\sigma^f(\omega,T)$ is used\cite{Bul98}.

Single-particle Green's function $G_\sigma(\omega,T)$ for the conduction electrons can be written as\cite{Hew93}:
\begin{equation}
G_\sigma (\omega ,T) = G_\sigma^0 (\omega ,T) + G_\sigma^0 (\omega ,T) c_{imp}T_\sigma (\omega ,T)G_\sigma^0 (\omega ,T),
\end{equation}
where $G_\sigma^0(\omega,T)$ is the single-particle Green's function for the non-interacting conduction electrons. 
The single impurity $T$-matrix $T_\sigma(\omega,T)$ is given by $|V_{hyb}(\omega)|^2G_\sigma^f (\omega ,T)$\cite{Hew93}. 
For a small impurity concentration $c_{imp}$, $G_\sigma(\omega,T)$ is determined by the Dyson equation:
\begin{equation}
G_\sigma (\omega ,T) = G_\sigma^0 (\omega ,T) + G_\sigma^0 (\omega ,T) c_{imp}T_\sigma (\omega ,T)G_\sigma (\omega ,T),
\end{equation}
which is valid to the first order of $c_{imp}$. In this approximation, the self-energy, $\Sigma_\sigma(\omega,T)$, 
for the conduction electrons is equal to $c_{imp}T_\sigma(\omega,T)$. Then, the relaxation time can be obtained by taking 
the imaginary part of $\Sigma_\sigma(\omega,T)$. By substituting $A_\sigma^f(\omega,T)$ for $T_\sigma(\omega,T)$, 
the expression of the relaxation time can be written as:
\begin{equation}
\frac{1}{\tau_\sigma (\omega ,T)} = \frac{2\pi c_{imp} |V_{hyb}(\omega)|^2}{\hbar}A_\sigma^f (\omega ,T).
\label{totalrate}
\end{equation}
Finally, based on the Boltzmann transport theory and using $\tau_\sigma(\omega,T)$ obtained above, 
the resistivity (sum over two spin) can be written in the form:
\begin{equation}
\rho_m(T) = \rho_0 [\int(-\frac{\partial f(\omega)}{\partial \omega})\frac{|\omega +\mu|d\omega}{|V_{hyb}(\omega)|^2 A_\sigma^f (\omega ,T)}]^{-1}
\label{resis}
\end{equation}
where $\rho_0 = 4\pi^2 c_{imp}\hbar/e^2$ and $f(\omega)$ is the Fermi-Dirac distribution function.

\subsection{Dephasing rate calculation}
An electron could interact with another electron through an inelastic scattering event. This scattering would change 
its energy and hence the evolution of its phase. Consequently, the phase of the electron wave would suffer some dephasing. 
Therefore, the dephasing rate can be defined as the inelastic scattering rate. The total scattering rate has 
alrealy been given in Eq. \eqref{totalrate}. The elastic scattering rate can be derived from 
the Fermi golden rule, and written as:
\begin{equation}
\frac{1}{\tau_\sigma^{(ela)} (\omega ,T)} = \frac{2c_{imp} |V_{hyb}(\omega)|^2}{\hbar}\Gamma(\omega)|G_\sigma^f (\omega ,T)|^2.
\end{equation}
Finally, the dephasing rate can be obtained as the difference between the total scattering 
rate and the elastic scattering. Therefore, the $\omega$-resolved dephasing rate can be written as\cite{Zar04}:
\begin{equation}
\begin{split}
\frac{1}{\tau_{\sigma}^{\phi}(\omega ,T)} = \frac{2c_{imp} |V_{hyb}(\omega)|^2}{\hbar}\ \ \ \ \ \ \ \ \ \ \ \ \ \ \ \ \ \ \ \ \ \ \ \\
\times [\pi A_\sigma^f (\omega ,T)-\Gamma(\omega)|G_\sigma^f (\omega ,T)|^2].
\end{split}
\end{equation}

In experiments, measurement of the resistivity correction to the weak localization effect can be used to determine 
the dephasing rate\cite{Alt85}. To compare to experiments, one should calculate the total dephasing rate $\gamma_m(T)$, 
which for two-dimensional structures is given by an integral over the $\omega$-resolved dephasing rate as\cite{Mic06}:
\begin{equation}
\gamma^\phi_m(T) = \frac{1}{\tau}\textrm{exp}[\int (-\frac{\partial f(\omega)}{\partial\omega})\textrm{ln}\frac{\tau}{\tau_\sigma^{\phi}(\omega,T)}d\omega],
\label{dephasingrate}
\end{equation}
where $\tau$ is the unit time.

\section{Results and discussion}

\begin{figure}
\includegraphics[width= 8cm]{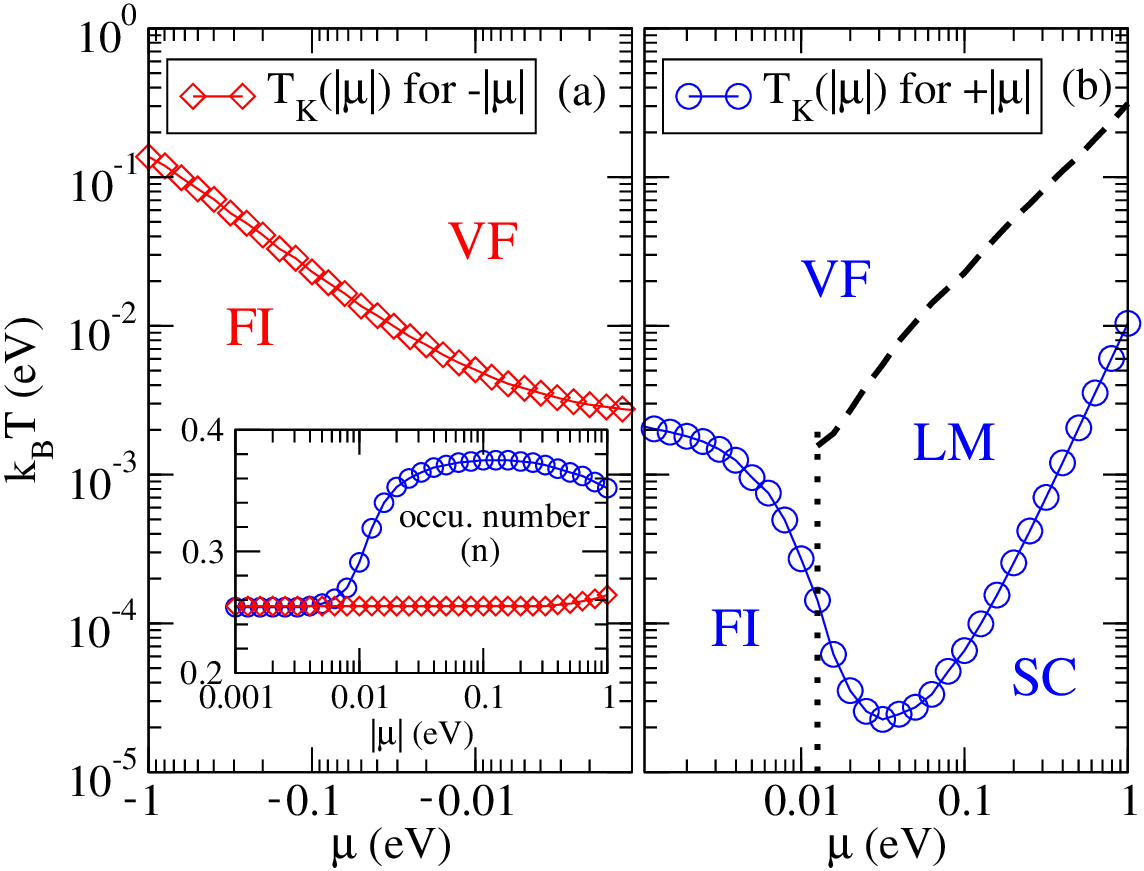}\\
\caption{(color online) Kondo temperature $T_K$ as a function of chemical potential $\mu$. $T_K$ is defined 
by equation $S_{imp}/k_B=(1/2)\textrm{ln}2$.
VF denotes the valence-fluctuation regime, LM the local-moment regime, FI the frozen-impurity regime, and SC the 
strong-coupling regime. In (a), the zero-temperature occupation number ($n$) of the impurity spin-up (spin-down) state
is displayed as a function of the magnitude of chemical potential ($|\mu|$) in the inset.
In (b), the dashed line, determined by equation $S_{imp}/k_B=1.2\textrm{ln}2$,
indicates the approximate boundary between the VF and LM regimes;
The dot line indicates the approximate boundary between the FI and SC regimes (see Sec. IIIC). 
}
\label{TK}
\end{figure}

Before presenting the calculated transport properties, let us first examine the phase diagram
of the Kondo effect in graphene. In the pure graphene (i.e., $\mu$ = 0), there are two stable fixed points, 
i.e., the local-moment (LM) and frozen-impurity (FI) fixed points\cite{Kri80}.
The system would flow to either the LM or FI regime in the low-temperature limit, depending critically on 
the $\Gamma_0$ value.
When the Fermi level is raised or lowered slightly, however, the LM fixed point would become unstable.
Also, there would be another stable fixed point, namely, the strong-coupling (SC) fixed point, when $\mu$ is positive.
Nevertheless, although the system with $\Gamma_0$ being smaller than the critical value
could enter the SC regime\cite{Kri80}, $T_K$ could be very low\cite{Voj10}, 
which would not be  experimentally accessible. Therefore, we only investigate the cases 
with $\Gamma_0$ being larger than the critical value ($\Gamma_c = 0.953$ eV).

Note that in the recent literature\cite{Fri13,Jac10,Kan12,Voj10}, the FI and SC fixed points are generally refered 
to as the asymmetric SC point. We believe that this is at least not precise since,
as we demonstrate below, the spectral function and transport properties are 
quite different in the FI and SC regimes. Therefore, here we consider the FI and SC fixed points (regimes)
as two different fixed points (regimes). Nevertheless, in this paper, we loosely use the Kondo temperature
$T_K$ to denote the temperature where the system would enter either the SC or FI regime from either the LM
or VF regime for simplicity. Strictly speaking, one should use another symbol (e.g., $T^*$) to denote the
transition from the VF to FI regime since there is no Kondo effect in this crossover.

Figure \ref{TK} shows the dependence of the Kondo temperature and also the impurity state 
occupation number ($n$) on the chemical potential.
Here, $T_K$ is defined by equation $S_{imp}/k_B=(1/2)\textrm{ln}2$ where $S_{imp}$ is the impurity entropy and
$k_B$ is the Boltzmann constant, respectively. 
It is clear that there is a pronounced particle-hole asymmetry in the graphene Kondo systems\cite{Voj10}.
As mentioned in Sec. II.A, the graphene Kondo systems with the types of impurities we consider here,
are a mixed valence system. 
Therefore, as one might expect from the conventional Anderson impurity model, there is no LM fixed point 
and only one crossover, namely, the crossover from the 
valence-fluctuation (VF) to FI regime\cite{Kri80} for the negative $\mu$ values, as shown in Fig. \ref{TK}(a). 
When the Fermi level is lowered below the Dirac point, the density of states at the Fermi level 
increases and so is the scattering rate. Therefore, $T_K$ would increase with $|\mu|$ [Fig. \ref{TK}(a)].
In the positive $\mu$ case, in contrast, $T_K$ initially decreases with $|\mu|$ and then reaches a minimum 
near $\mu$ = 0.04 eV. As the Fermi level is further raized, $T_K$ would increase monotonically with $|\mu|$. 
We could attribute this interesting behavior to the occurrence of the LM fixed point when $\mu > 0.0126$ eV,
which is also evident from the sharp increase of the occupation number of the impurity state around $\mu \approx 0.0126$ eV
[See the inset in Fig. \ref{TK}(a)].
Therefore, the system would enter the SC regime at lower temperatures from the LM regime (i.e., the Kondo effect) 
rather than the VF regime [Fig. \ref{TK}(b)]. 

\subsection{Resistivity vs. temperature}

\begin{figure}
\includegraphics[width= 8cm]{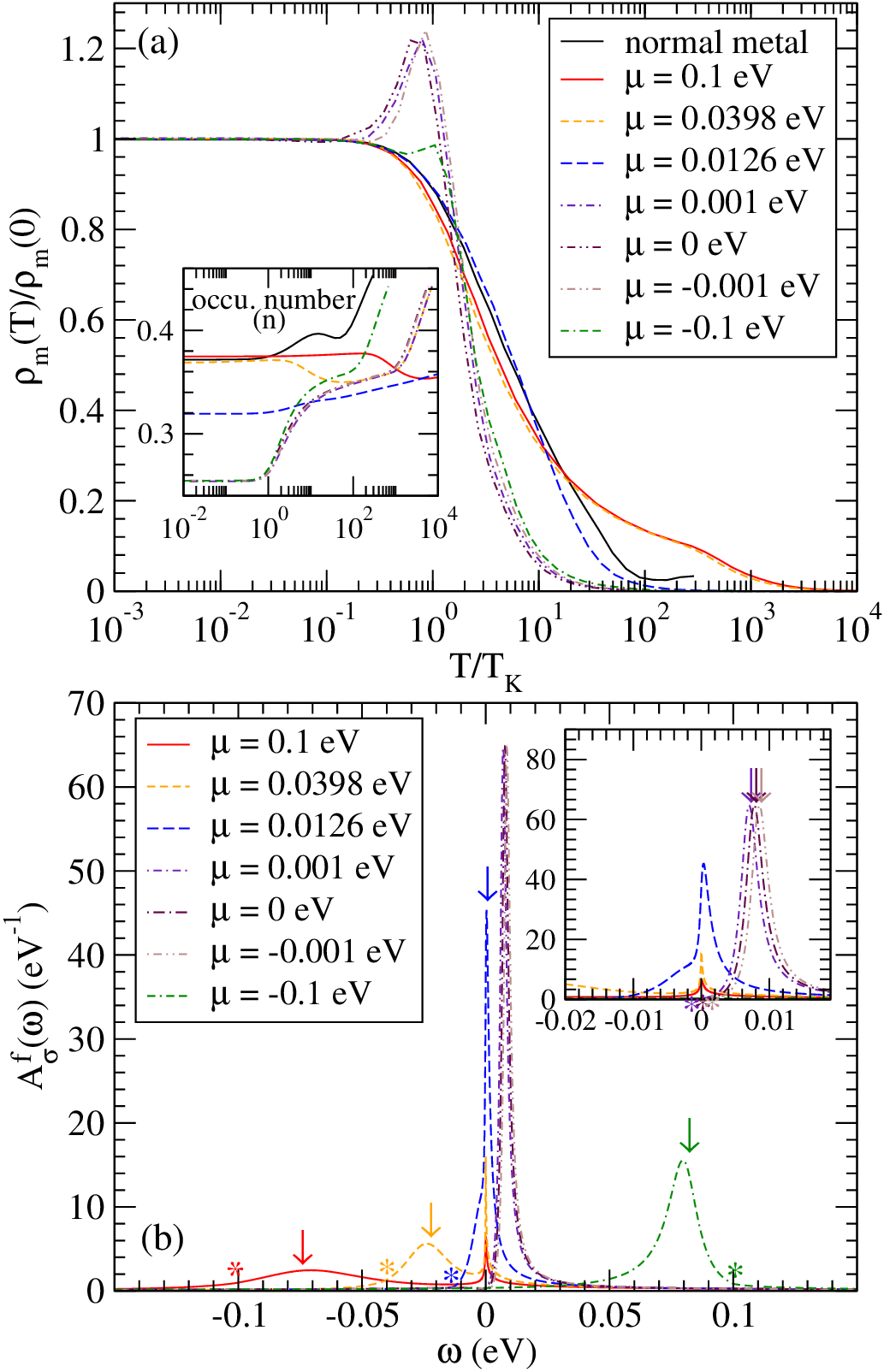}
\caption{(color online) (a) Resistivity as a function of temperature for different chemical potentials. 
Occupation number ($n$) of the impurity spin-up (spin-down) state as a function
of temperature for different chemical potentials is displayed in the inset in (a). 
(b) Impurity spectral function at zero temperature under different 
chemical potentials $\mu$. $\omega$ = 0 is 
the Fermi level. The stars indicate the positions of the Dirac point; 
the arrows denote the positions of the effective (renormalized) impurity level ($\tilde{\epsilon_d}$). 
}
\label{Resistivity}
\end{figure}

Figure \ref{Resistivity} shows the $T$-dependence of the resistivity 
and also the spectral function at $T = 0$ under different chemical potentials.
In the present calculations, we first set $s(x) =$ 1, 
and the scattering rate becomes the power-law of $r =$ 1. This means that for the case 
of carbon vacancies the higher order terms in the scattering 
rate are neglected, and that for the case of adatoms only the hybridization 
of the impurity with conduction electrons below the impurity is considered. 
In general, all the resistivity-versus-$T$ curves in the graphene Kondo systems, 
especially that for $\mu = 0.0126$ eV, look similar to that of the usual Kondo effect\cite{Nor}.
As the systems go through the crossover region from the high-$T$ VF regime to low-$T$ FI (or SC) regime, 
the resistivity increases steeply and eventually saturates to the zero-temperature resistivity [$\rho_m(0)$] near $T_K$. 

\begin{figure}
\includegraphics[width= 8cm]{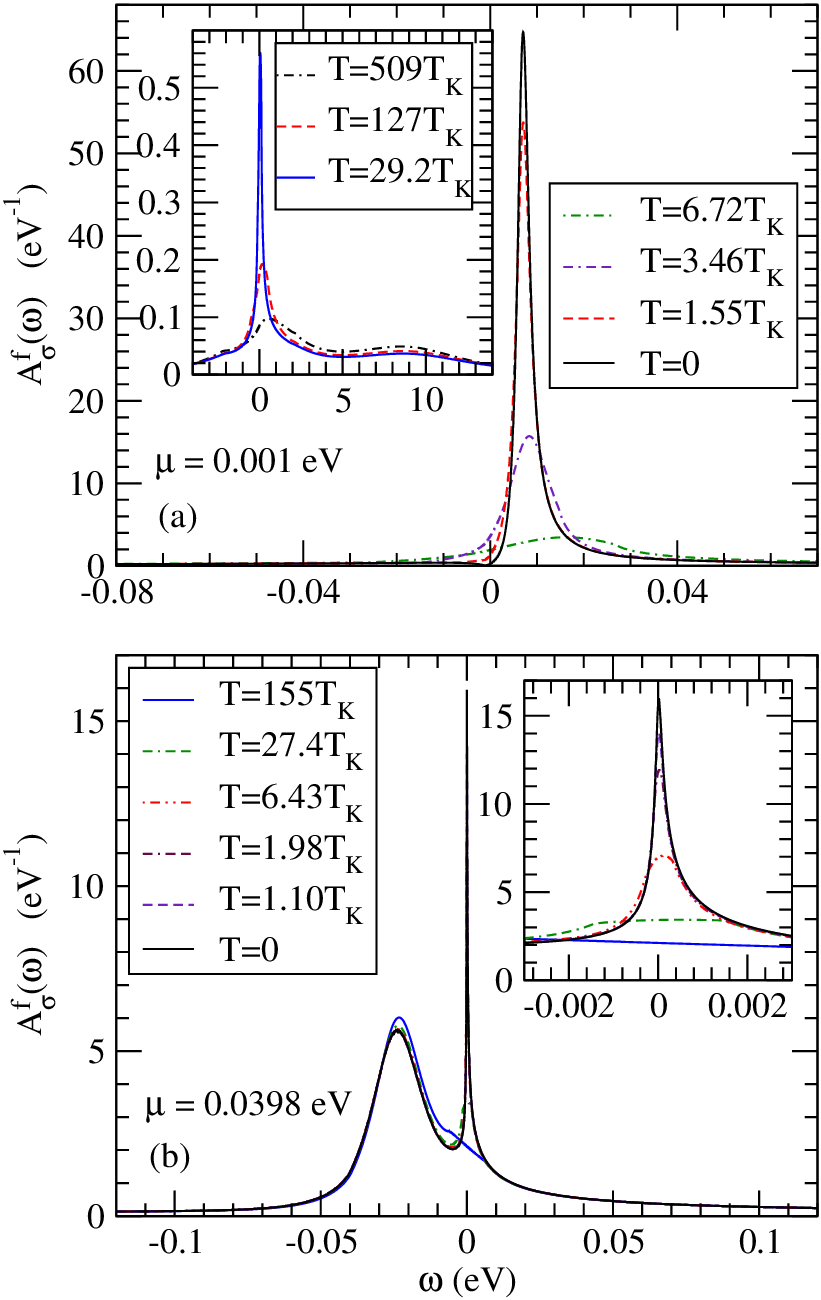}\\
\caption{(color online) Impurity spectral function for (a) $\mu$ = 0.001 eV and (b) $\mu$ = 0.0398 eV
at different temperatures. $\omega$ = 0 is the Fermi level.}
\label{Fig3}
\end{figure}

Nevertheless, a closer examination of Fig. \ref{Resistivity}(a) shows that many resistivity-versus-$T$ 
curves exhibit certain prominent features that are distinctly different from that in normal metals.
In particular, the resistivity-versus-$T$ curves can be grouped into three cases depending on their
chemical potentials [Fig. \ref{Resistivity}(a)]. The first case includes the systems with 
negative $\mu$ and also very small $|\mu|$ (e.g., $\mu =$ 0 and $\pm$0.001 eV) values.
The resistivity curves in this case  have a pronounced peak near $T_K$, especially for very 
small $|\mu|$ values. When the temperature is lowered from about 100$T_K$, the resistivity first rises rapidly 
to that above $\rho_m(0)$, but then drops quickly to $\rho_m(0)$ as temperature further decreases. 
This interesting feature is caused by the fact that the resonance of the impurity state is 
located at about 10 meV above the Fermi level [see the inset in Fig. \ref{Resistivity}(b)]. 
Therefore, as temperature decreases, the resonance peak becomes narrower and thus moves away from the Fermi level 
[see the inset in Fig. \ref{Fig3}(a)]. Furthermore, as $\mu$ becomes more negative, the peak would move further above
the Fermi level [see, e.g., the curve for $\mu = -0.1$ eV in Fig. \ref{Resistivity}(b)] and consequently,
the pronounced peak near $T_K$ would be reduced. Interestingly, there is no Kondo resonance at the Fermi level
in the impurity spectral function at low temperatures in this case [see, e.g., Fig. \ref{Fig3}(a)]. 
Therefore, the systems are driven from the high-$T$ VF regime to the low-$T$ FI regime rather than the SC regime.
Nevertheless, Fig. \ref{Resistivity}(a) shows that, in the crossover from the VF to the FI regime, 
the resistivity curves in this case approach their saturation value in the same manner.
However, the increase of the resistivity as temperature decreases, is
much more rapid [as $1.68 (T_K/T)^{3/2}$, as determined by a fitting to the resistivity curves]
than that in the conventional Kondo effect [Figure \ref{Resistivity}(a)].

The case of large positive $\mu$ values ($\mu > 0.0126$ eV) belongs to the second one.
A common feature in this case is that apart from the resonance peak due to the
impurity state, the sharp Kondo resonance appears at the Fermi level in
the spectral function [see, e.g., the curves for $\mu = 0.0398$ and 0.10 eV in Fig. \ref{Resistivity}(b)]. 
As a result, the resistivity curve can be divided into two parts. 
The first part with $T \ge 10T_K$, 
is mainly caused by the scattering of the local moment. The other part
with $T < 10T_K$, is due to the strong coupling between the impurity spin and conduction electrons.
Figure \ref{Fig3}(b) shows that in the spectral function for $\mu = 0.0398$ eV, a wide peak centered
at the effective impurity level $\tilde{\epsilon}_d$ = -0.021 eV (also the Dirac point) and extended across the Fermi level,
forms at high temperatures. The sharp Kondo resonance at the Fermi level shows up as
temperature further decreases. The scattering of the local moment gives rise to
the broad resistivity shoulder above 10$T_K$ [Fig. \ref{Resistivity}(a)]. The Kondo resonance then 
causes the resistivity to rises rapidly, like the usual Kondo effect, as the system enters the SC regime.
This interesting shoulder-peak feature can be attributed to the occurrence of the 
additional LM fixed point [Fig. \ref{TK}(b)], 
as discussed already in the beginning of this section. 

The third case includes the systems with a positive $\mu$ being close to $\mu$ = 0.0126 eV. 
In this case, the impurity state peak is located at the Fermi level [Fig. \ref{Resistivity}(b)] 
and hence is merged with the Kondo resonance. There is no broad resistivity shoulder above 10$T_K$. 
The resistivity curves look almost identical to that of conventional Kondo 
systems [Fig. \ref{Resistivity}(a)].

\subsection{Renormalization of the impurity level}

In general, the impurity level in the Anderson model would be renormalized due to the Coulomb 
correlations\cite{Bul98}. As a result, the effective impurity level $\tilde{\epsilon}_d$ in the pure graphene 
would be shifted towards the Fermi level. The $U/|\epsilon_d|$ ratio and the $\Gamma_0$ value would determine 
whether the $\tilde{\epsilon}_d$ is below or above the Fermi level. With $\Gamma_0$ being smaller 
than the critical value $\Gamma_c$, $\tilde{\epsilon}_d$ would be located below the Fermi level. 
The system would flow to the LM regime in this case. Furthermore, since the density of states 
at the Fermi level is zero, the impurity magnetic moment could not be completely screened. Consequently, 
the system would stay in the LM regime at low temperatures. However, in the asymmetric Anderson 
model ($U\neq 2|\epsilon_d|$) with $\Gamma_0>\Gamma_c$, $\tilde{\epsilon}_d$ would be located 
above the Fermi level, and the system would flow to the FI regime.

\begin{figure}
\includegraphics[width= 8cm]{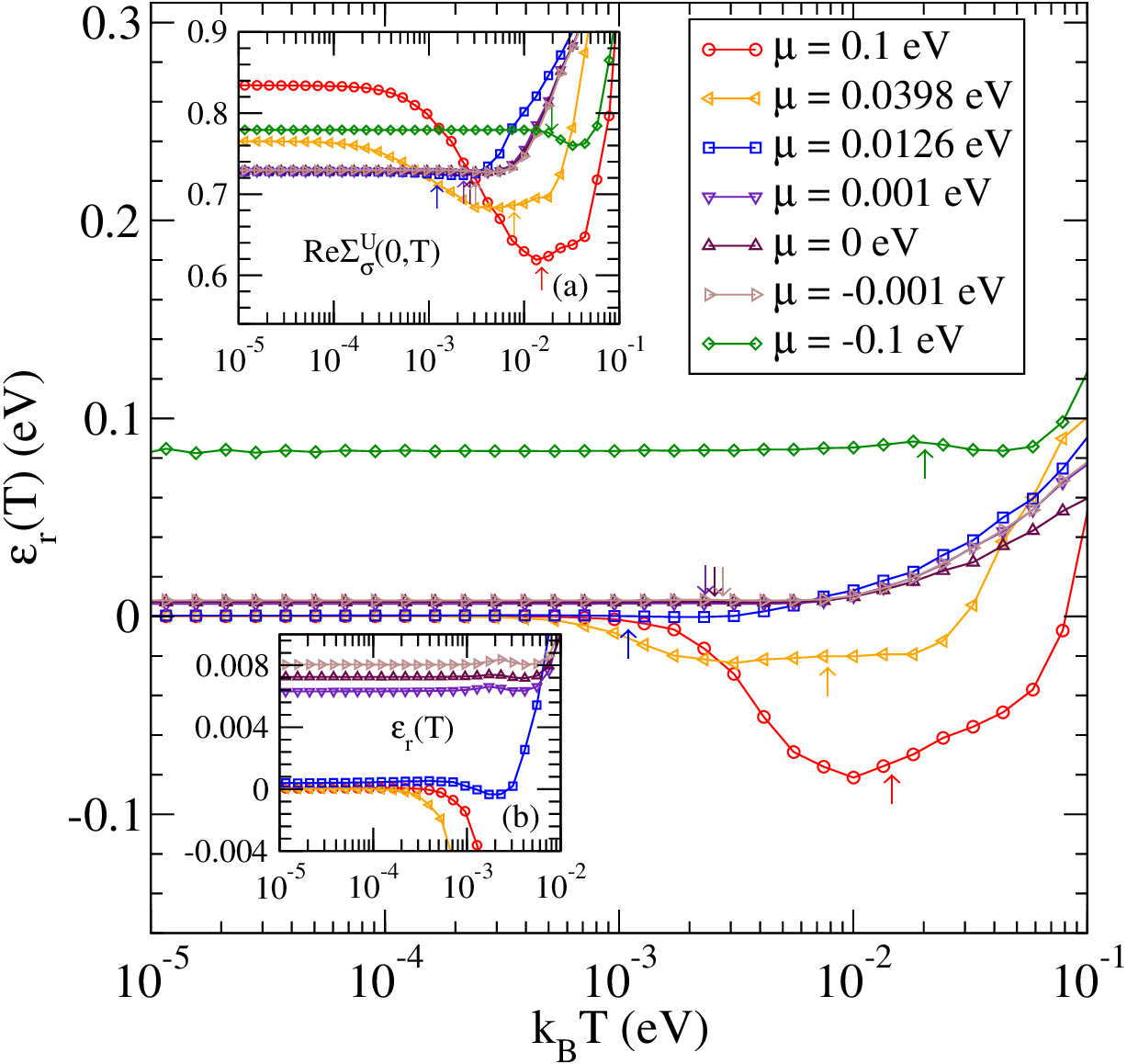}\\
\caption{(color online) The energy of the resonant peak [$\epsilon_r(T)$] in the impurity 
spectral function for different chemical potentials as a function of temperature.
The real part of the self-energy due to the Coulomb correlations [Re$\Sigma^U_\sigma(0,T)$] 
for different chemical potential as a function of temperaure is displayed in inset (a).
The arrows indicate the temperatures ($T_b$) at which $\epsilon_r(T_b)$ equals to the energy of
the effective impurity level ($\tilde{\epsilon}_d$). 
}
\label{Ed}
\end{figure}

To evaluate the effective impurity level and also to see how the position of the resonant 
peak ($\epsilon_r$) in the impurity spectral function evolves as temperature varies, 
let us rewrite the impurity Green's function as 
\begin{equation}
G_\sigma^f(\omega,T)= \frac{1}{(\omega+i\delta)-\epsilon_d-\Sigma^U_\sigma(\omega,T)-\Delta_\sigma(\omega)},
\end{equation}
where $\delta \rightarrow 0^+$, $\Sigma^U_\sigma(\omega,T)$ is the self-energy due to the Coulomb correlations 
and $\Delta_\sigma(\omega)$, given by
\begin{equation}
\Delta_\sigma(\omega) = \int_{-D}^D\frac{\Gamma(\omega)/\pi}{(\omega+i\delta)-\omega'}d\omega',
\end{equation} 
is the self-energy caused by the asymmetry of $\Gamma(\omega)$.
By expanding both $\Sigma^U_\sigma(\omega,T)$ and $\Delta_\sigma(\omega)$ in powers of $\omega$,
$G_\sigma^f(\omega,T)$ can be rewritten as:
\begin{equation}
G_\sigma^f(\omega,T)= \frac{z}{(\omega+i\delta)-\epsilon_r(T)-i\textrm{Im}\tilde{\Delta}_{\sigma}},
\end{equation}
where $z^{-1}=1-Re[\partial\Sigma^U_\sigma(\omega,T)/\partial\omega]_{\omega=0}-[\partial\Delta_\sigma(\omega)/\partial\omega]_{\omega=0}$,
$\tilde{\Delta}_{\sigma}=z\Delta_\sigma(0)$, and $\epsilon_r(T)=z[\epsilon_d+\textrm{Re}\Sigma^U_\sigma(0,T)+\textrm{Re}\Delta_\sigma(0)]$
is the energy position of the resonance. Note that the meaning of $\epsilon_r(T)$ differs in the different regimes. 
In particular, in the FI or LM regime, $\epsilon_r(T)$ is the energy of the effective impurity level $\tilde{\epsilon}_d$,
whereas in the SC regime, it is the position of the Kondo resonance. 

Figure \ref{Ed} shows $\epsilon_r(T)$ and Re$\Sigma^U_\sigma(0,T)$ for different chemical potentials 
as a function of temperature. As the temperature gets lowered from the high-$T$ 
VF regime, $\epsilon_r(T)$ would decrease monotonically until the boundary between the VF and 
FI (LM) regimes for $\mu < 0.0126$ eV ($\mu > 0.0126$ eV) (Fig. \ref{TK}). At the boundary ($T = T_b$), the impurity
level forms and $\epsilon_r(T)$ becomes equal to the energy of the effective impurity 
level $\tilde{\epsilon}_d$ [i.e., $\tilde{\epsilon}_d=\epsilon_r(T_b)$], 
as indicated by the arrows in Fig. \ref{Resistivity}(b) and Fig. \ref{Ed}. 
Clearly, the lowering of $\epsilon_r(T)$ is caused by the decrease of Re$\Sigma^U_\sigma(0,T)$ due to
the increasing strength of the weak interaction between the impurity and conduction electrons [see inset (a) in Fig. \ref{Ed}].
Interestingly, at $T = T_b$, Re$\Sigma^U_\sigma(0,T) \approx -(\epsilon_d+\mu)$ [see inset(a) in Fig. \ref{Ed}].

In the case of $\mu < 0.0126$ eV, Re$\Sigma^U_\sigma(0,T)$ and hence $\epsilon_r(T)$ 
would then remain unchanged when temperature is further lowered.
For very small $|\mu|$ values (e.g., $\mu$ = 0, $\pm$0.001 eV), $\tilde{\epsilon_d}$ is located right above 
the Fermi level [see Fig. \ref{Resistivity}(b) and inset(b) in Fig. \ref{Ed}]. 
For the negative $\mu$ (e.g., $\mu$ = -0.1 eV), $\tilde{\epsilon_d}$ would move further above the Fermi level.
Therefore, the systems would flow to the FI regime since the impurity state is unbound ($\tilde{\epsilon_d}>0$).
In contrast, for $\mu > 0.0126$ eV (e.g., $\mu$ = 0.0398 or 0.1 eV), 
$\tilde{\epsilon_d} < 0$ and the bound impurity state forms near $T_b$. 
Therefore, the system would first flow to the LM regime.
As temperature further decreases, the impurity state would then interact very strongly with the conduction electrons,
resulting in the pronounced increase of Re$\Sigma^U_\sigma(0,T)$ [see inset (a) in Fig. \ref{Ed}].
Consequently, $\epsilon_r(T)$ would be renormalized to the Fermi level, 
and the sharp Kondo resonance at the Fermi level would develop [Fig. \ref{Fig3}(b)], 
as the systems move from the LM regime to the SC regime [see Fig. \ref{TK} (b) and Fig. \ref{Ed}].
The local magnetic moment would then be screened at low temperatures.
Clearly, $\mu$ = 0.0126 eV is the critical $\mu$ value that separates the FI and SC regimes [see Fig. \ref{TK}(b)].
In this case, $\tilde{\epsilon}_d = 0$, i.e., the effective impurity level is located right at the Fermi level.

\subsection{Dephasing rate vs. temperature}

\begin{figure}
\includegraphics[width= 8cm]{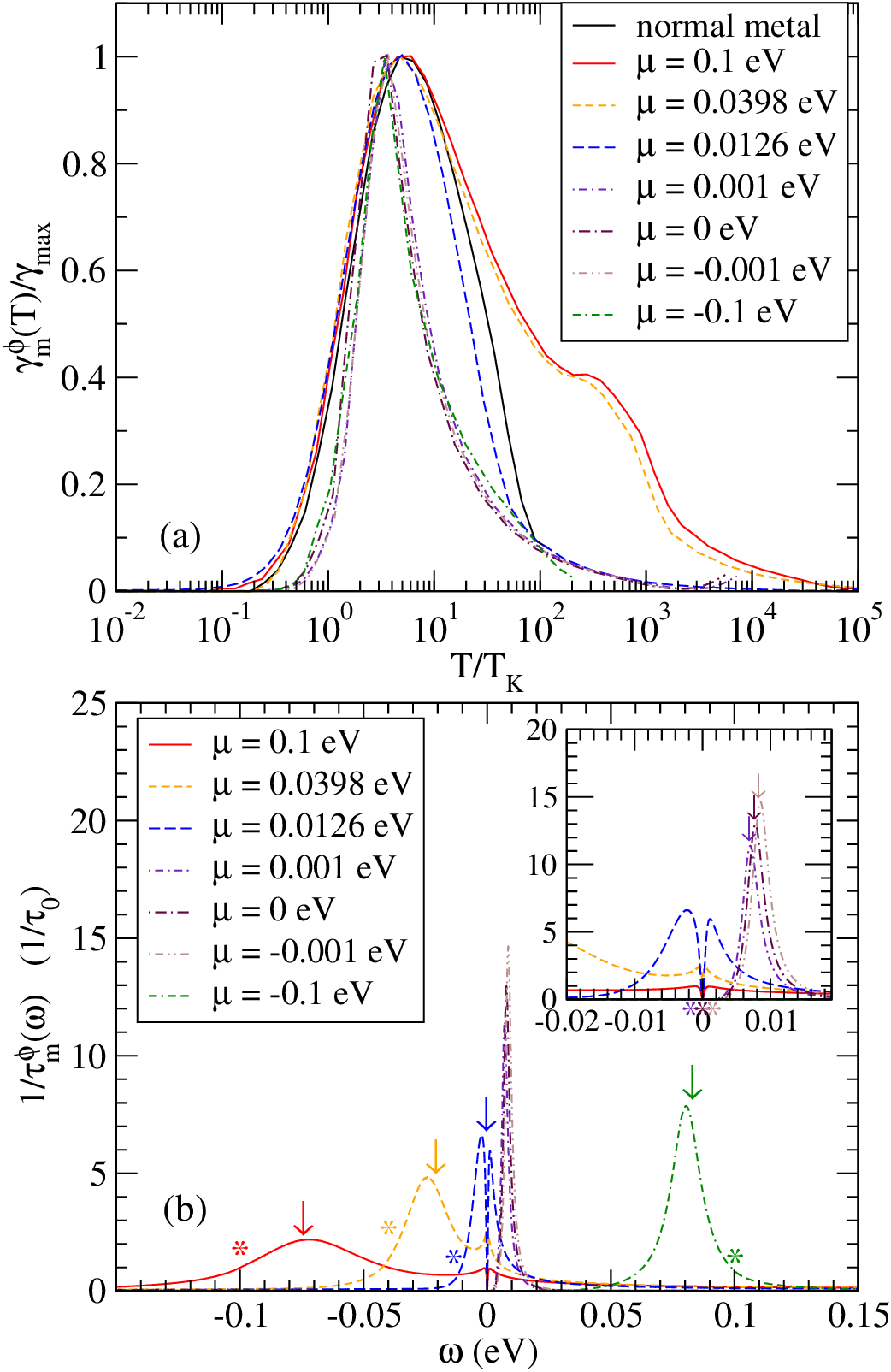}\\
\caption{(color online) (a) Total dephasing rates as a function of temperature for different chemical potentials.
$\gamma_{max}$ is the maximal dephasing rate. 
(b) Zero temperature $\omega$-resolved dephasing rates for different chemical potentials $\mu$. 
$1/\tau_0$ = $2c_{imp}\hbar v_F^2/\Omega_0|\mu_0|$ with $\mu_0$ = 1 eV. 
$\gamma_{max}$ is the maximal dephasing rate. $\omega$ = 0 is 
the Fermi level. The stars indicate the positions of the Dirac point;
the arrows denote the positions of the effective impurity level ($\tilde{\epsilon_d}$).
}
\label{DephasingRate}
\end{figure}

\begin{figure}
\includegraphics[width= 8cm]{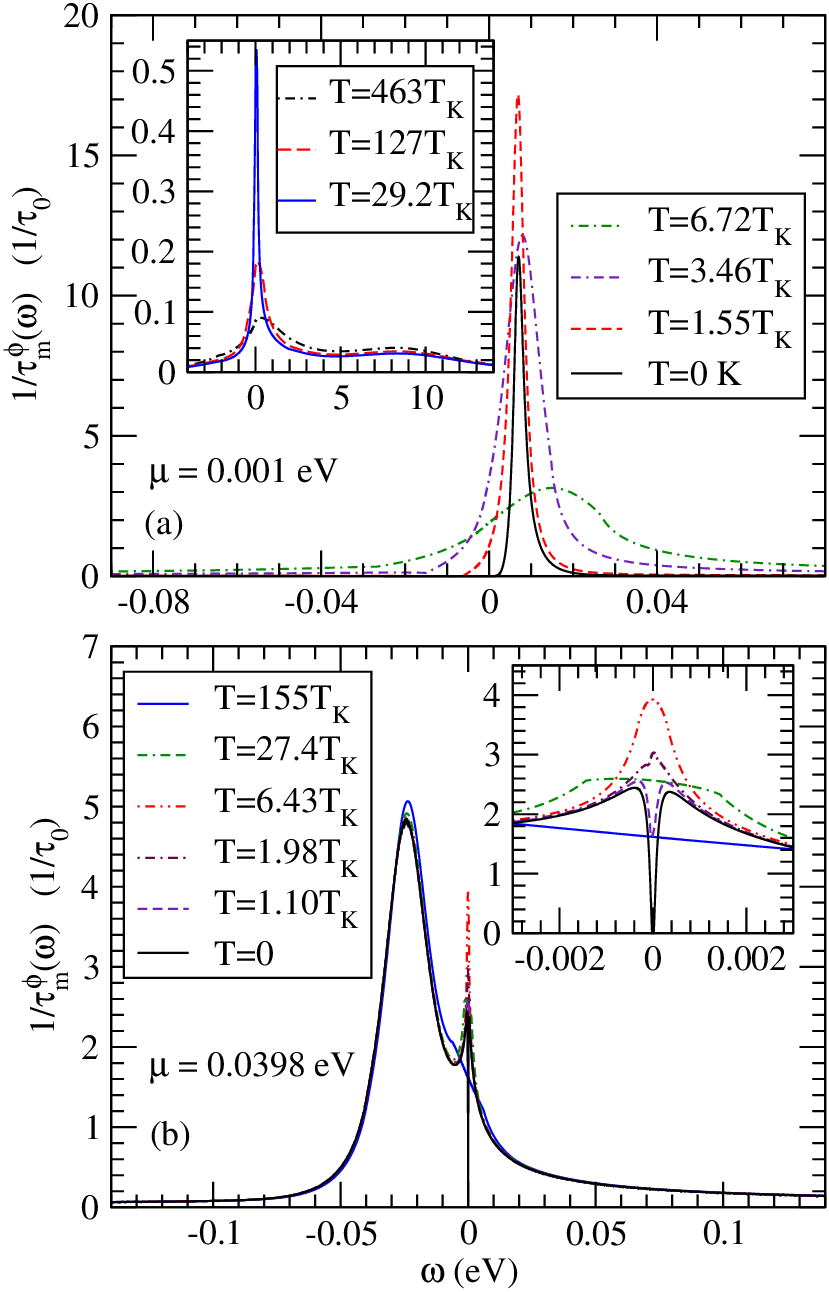}\\
\caption{(color online) $\omega$-resolved dephasing rates for (a) $\mu$ = 0.001 eV and (b) $\mu$ = 0.0398 eV
at different temperatures. $1/\tau_0$ = $2c_{imp}\hbar v_F^2/\Omega_0|\mu_0|$ with $\mu_0$ = 1 eV. 
$\omega$ = 0 is the Fermi level.} 
\label{Fig6}
\end{figure}

The calculated dephasing rates as a function of temperature
for different chemical potentials are displayed in Fig. \ref{DephasingRate}(a).
To help understand the calculated dephasing rates, we show the $\omega$-resolved dephasing rates at zero temperature
for different chemical potentials in Fig. \ref{DephasingRate}(b). For the same purpose, we also display the $\omega$-resolved 
dephasing rates at different temperatures for $\mu$ = 0.001 and 0.0398 eV in Fig. \ref{Fig6}. 
The zero-temperature $\omega$-resolved dephasing rates are very similar to 
the corresponding impurity spectral functions [Fig. \ref{Resistivity}(b)]
except at the Fermi level where the $\omega$-resolved dephasing rates are always zero in the $T = 0$ limit. 
In general, however, the curves of the dephasing rate as a function of temperature look quite different from that 
of the resistivity [see Figs. \ref{Resistivity}(a) and \ref{DephasingRate}(a)]. Figure \ref{DephasingRate}(a) shows that
all the dephasing rate curves constitute mainly a prominent peak in the crossover region. 
When the system enters the crossover region, the dephasing rate increases rapidly as temperature decreases,
and reaches the maximum near the center of the crossover region. However, as temperature further decreases, the
system would enter the SC (FI) regime for $\mu > 0.0126$ eV ($\mu < 0.0126$ eV).
Consequently, the dephasing rate drops sharply and finally vanishes in the $T = 0$ limit.

Figure \ref{DephasingRate}(a) shows that, as for the resistivity curves, the dephasing rate curves can be
grouped into three cases. 
In this first case ($\mu < 0.0126$ eV), the dephasing rate curves have a narrower peak compared 
to that of the conventional Kondo effect. Indeed, the dephasing rate begins to increase at a lower $T/T_K$
than that in normal metals and also in the systems with a large positive $\mu$ (e.g., $\mu = 0.0398$ or $0.1$ eV).
Furthermore, the increasing slope is steep and a fitting to the dephasing rate curve gives a rising function
of $\sim4.6(T_K/T)$. In the lower temperature side of the peak, the dephasing rate drops more quickly 
than that in normal metals and also for $\mu \ge 0.0126$.
A fitting to the dephasing rate curves gives a droping function of $\sim0.10(T/T_K)^3$, being distinctly different
from the $(T/T_K)^2$ behavior obtained using Fermi liquid theory for normal metals~\cite{Hew93}
This interesting behavior may be attributed to the fact that the effective impurity level is located 
above the Fermi level ($\tilde{\epsilon_d}>0$) and thus the Fermi level is not in the range of the impurity 
state peak in the low-$T$ limit [Figs. \ref{Resistivity}(b), \ref{DephasingRate}(b) and \ref{Fig6}(a)]. 
Therefore, in the crossover from the VF to the FI regime, the dephasing rate, after reaching the maximum, 
would drop sharply since the width of the peak rapidly becomes narrower with the decreasing temperature 
[see Fig. \ref{Fig6}(a)]. All scatterings, including the inelastic scattering, would be suppresed 
in the $T= 0$ limit.

In the second case ($\mu > 0.0126$ eV), although the dephasing rate curve is almost identical to the curve 
of the conventional Kondo effect below the peak temperature ($T_m$), the dephasing rate curve differs
significantly above $T_m$ [Fig. \ref{DephasingRate}(a)]. In particular, the dephasing rate curve has a 
pronounced shoulder above $100T_K$. This is caused by the occurrence of the LM fixed point in this case [Fig. \ref{TK}]. 
In the LM regime, the dephasing rate first rises well above $100T_K$ due to the inelastic scattering 
of the local moment, which gives rise to a broad peak covering the Fermi level [Fig. \ref{DephasingRate}(b)].
This results in a wide shoulder above 100$T_K$. In the crossover from the LM to the SC regime, 
the dephasing rate then rises more quickly to the maximum since the impurity now becomes 
entangled with the conduction electrons. This is illustrated for $\mu > 0.0398$ eV in the inset in Fig. \ref{Fig6}(b) 
where it is seen that the Kondo resonance grows rapidly as $T_K$ is approached from above. 
As temperature further decreases, however, the dephasing rate drops sharply and eventially vanishes 
because the Kondo resonance starts to decrease and eventually drops to zero [see the inset in Fig. \ref{Fig6}(b)]
since the impurity magnetic moment is now totally screened.

In the third case (i.e., $\mu \approx 0.0126$ eV), the dephasing rate curve is similar
to that for the conventional Kondo effect. In particular, the two dephasing rate curves 
are nearly identical in the low temperature side of the peak [Fig. \ref{DephasingRate}(a)].
This is because at this $\mu$ value, the Fermi level falls within the range of the 
impurity state peak in the spectral function [Figs. \ref{Resistivity}(b) and \ref{DephasingRate}(b)].
This causes the behavior of the dephasing rate to be similar to that of the conventional Kondo effect.

\subsection{Saturation resistivity}

\begin{figure}
\includegraphics[width= 8cm]{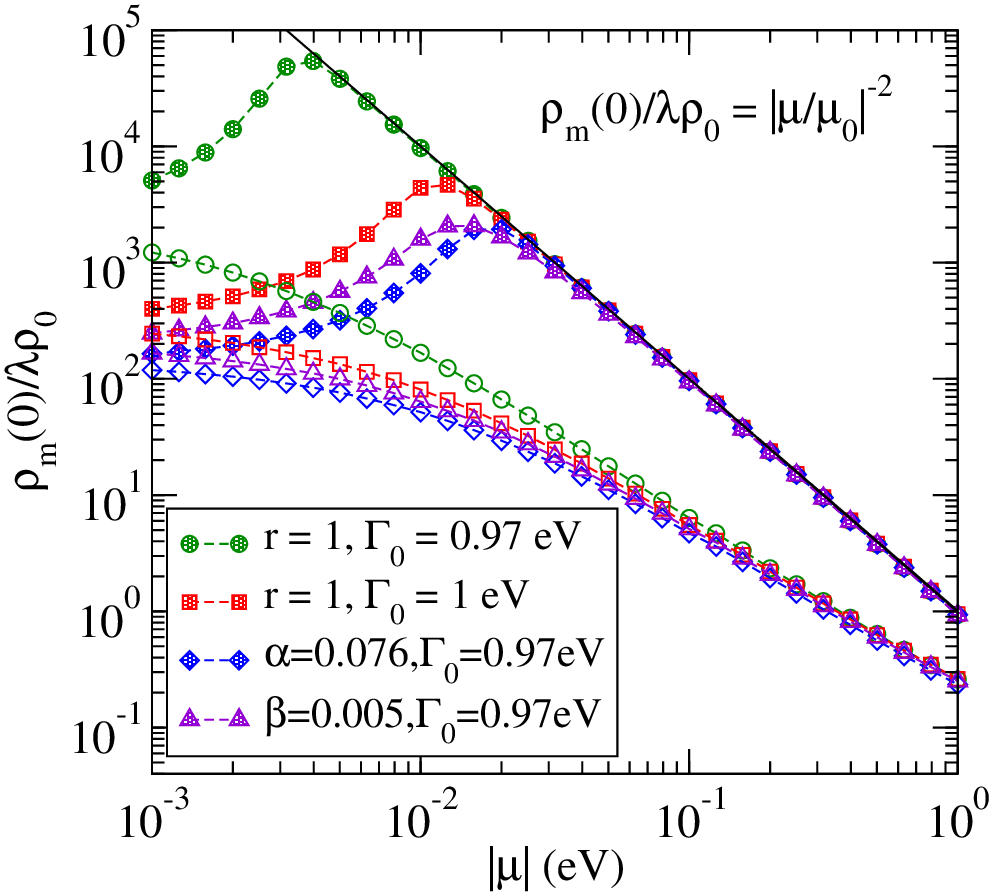}\\
\caption{(color online) Resistivity at zero temperature as a function of chemical potential. 
Filled and open symbols are for the positive and negative chemical potentials, respectively. 
Blue diamond lines are for the case of carbon vacancies, 
and purple up-triangle lines are for the case of magnetic adatoms. 
}
\label{Rvs.Vg}
\end{figure}  

As temperature tends to zero, $-(\partial f(\omega)/\partial \omega) \approx \delta (\omega)$, 
and hence the saturation (i.e., zero temperature) resistivity can be approximated as: 
\begin{equation}
\label{resistivity}
\rho_m(0)_{\mu \neq 0} = \frac{\rho_0|V_{hyb}(0)|^2 A_\sigma^f (0 ,0)}{|\mu|}.
\end{equation} 
For the $\mu$ = 0 case, the zero temperature resistivity should be calculated by using equation
\begin{equation}
\label{resistivity}
\rho_m(0)_{\mu = 0} = \frac{\rho_0|V_{hyb}(0)|^2 \Gamma_0/\pi D_{eff}}{[\epsilon_d+\Sigma^U_\sigma(0,0)]^2}.
\end{equation} 
The calculated zero temperature resistivity as a function of chemical potential is plotted in 
Fig. \ref{Rvs.Vg}. First of all, Fig. \ref{Rvs.Vg} shows that the $\rho_m(0)$ 
values for the negative $\mu$ values are more than 10 times smaller than that for the positive $\mu$ values,
further indicating a strong particle-hole asymmetry in the graphene Kondo systems. 
This can be explained as follows.
For the positive $\mu$ values ($\mu > 0.0126$ eV), as discussed in Sec. IIIB, $\tilde{\epsilon}_d < 0$ and 
thus the bound impurity state with the local moment is present. 
Consequently, the system would flow to the SC regime and the Kondo resonance would occur at the Fermi level
as temperature goes to zero. This Kondo effect results in a very strong scattering and
hence a very large resistivity at $T = 0$ K.
For the negative $\mu$ values, on the other hand, $\tilde{\epsilon}_d > 0$ and
thus the impurity state is unbound. As a result, the system would flow to the FI regime
and no Kondo resonance would develop at the Fermi level.
Therefore, the scattering would be much suppressed at low temperatures,
compared with the case of $\mu > 0.0126$ eV, resulting in a much smaller $T = 0$ resistivity.

Secondly, Fig. \ref{Rvs.Vg} also indicates that for large positive $\mu$ values 
($\mu > 0.0126$ eV), all the calculated resistivities at $T = 0$ K collapse 
onto the solid line determined by Eq. \eqref{R(V)}. 
In this case, as mentioned before, the system would flow to the SC regime and $\epsilon_r(T) \approx 0$
at low temperatures. Therefore, $A_\sigma^f(0,0)$ can be approximated as $1/(\pi \Gamma_0\frac{|\mu|}{D_{eff}})$. 
Plugging this into Eq. \eqref{resistivity}, the resistivity at $T = 0$ K can be written as
\begin{equation}
\label{R(V)}
\rho_m(0) = \lambda \rho_0 |\frac{\mu}{\mu_0}|^{-2},
\end{equation}
where $\lambda = \pi \Omega_0 |\mu_0|^{2}/(4\hbar^2 v_F^2)$ and $\mu_0$ = 1 eV. 
Clearly, this saturation value is independent of $\Gamma_0$, $\epsilon_d$ and $U$.
Therefore, the zero temperature resistivities collapse onto the solid line 
determined by Eq. \eqref{R(V)}. This is also a manifestation of the occurrence 
of the Kondo resonance in the graphene systems with a positive $\mu$ ( $\mu \ge 0.0126$ eV).

\subsection{Effect of higher order terms}

Finally, we consider the first higher order term. Figure \ref{Rvs.Vg} shows that the results 
for the cases of $\alpha \neq 0$ and $\beta \neq 0$ qualitatively agree with the case of the 
pure power-law of $r = 1$, albeit with a larger $\Gamma_0$. This is because for the case 
of carbon vacancies, total hybridization becomes larger when the higher order
terms are considered, resulting in a larger effective $\Gamma_0$.
For the case of magnetic adatoms, in the presence of the hybridization between the impurity and the 
conduction electrons on the next nearest neighbors, $\Gamma(\omega)$ becomes asymmetric,
giving rise to a larger Re$\Delta_\sigma(0)$. Therefore, the effective impurity level would rise.
As a result, the system needs a larger chemical potential to bring the effective impurity level
back to below the Fermi level, which has the same effect of a larger $\Gamma_0$.

\section{Comparison to experiments}

Recent experiments showed that the Kondo effect could be observed in graphene with point defects and vacancies\cite{Che11}. 
Two prominent features about the transport properties from the experiments were reported\cite{Che11}. 
Firstly, the Kondo effect was observed in graphene under a wide range of applied gate voltages (up to 
$|V_g| = \sim 50$ V, which is equivalent to $|\mu| = \sim 0.25$ eV).
Secondly, the normalized Kondo part of the resistivity appeared to be an universal function of $T/T_K(V_g)$, which
could fit well to that for the conventional Kondo effect obtained from much earlier NRG calculations\cite{Cos94}.
In principle, this could be explained as follows. With a large applied gate voltage, the Dirac point would 
lie well below the Fermi level. The two important properties of graphene, namely, zero density of states 
at the Dirac point and linear energy-dependence of the density of states, 
would hardly affect the transport properties of the system. Therefore, 
the system would behave like the conventional Kondo system. 
Indeed, as discussed before in Sec. IIIA, the shape of the resistivity-vs-$T$ curve 
for $\mu \approx 0.0126$ eV calculated here looks
very similar to that of the conventional Kondo effect [Fig. \ref{Resistivity}(a)].
However, as reported in Sec. IIIA, our calculated resistivity curves for other chemical potentials show
several unusual features which were not observed in the experiments\cite{Che11}. 

Very recently, the difficulties in uncovering the Kondo effect in graphene by resistivity measurements 
because of a possible similar contribution from the electron-electron interaction to the low temperature resistivity, 
were reported\cite{Job13}. On the other hand, Fig. \ref{DephasingRate}(a) shows that all the dephasing rates
in the graphene Kondo systems exhibit characteristic features that are distinctly different 
from that of the conventional Kondo effect. 
Therefore, we would suggest further measurements on the dephasing rate in graphene. 
With the distinct characteristics we have shown in Sec. III, 
the novel features of the resistivity and dephasing rate in graphene could be identified.

\section{Conclusions}
We have investigated the resistivity and dephasing rate in the Kondo regime due to magnetic impurities in graphene
under different gate voltages by NRG calculations.
We find that although the Anderson impurity model considered here is a mixed valence system,
it can be drived into either the Kondo [$\mu > \mu_c$] or 
mixed valency ($\mu \approx \mu_c$) or empty orbital ($\mu < \mu_c$) regime by a gate voltage,
thereby resulting in characteristic features in resistivity and dephasing rate in each regime.
In particular, in the case of $\mu < \mu_c$, the shapes of the resistivity (dephasing rate)
 curves are nearly identical. However, as temperature decreases, 
they start to increase to their maxima at a lower $T/T_K$ but more rapidly 
[as $(T_K/T)^{3/2}$] than in normal metals.
As $T$ further decreases, after reaching the maximum, the dephasing rate drops more quickly than in normal metals, 
behaving as $(T/T_K)^3$ instead of $(T/T_K)^2$. Furthermore, the resistivity has a pronounced 
peak above the saturation value near $T_K$.
In the case of $\mu > \mu_c$, in contrast, the resistivity curve has an additional broad shoulder above 10$T_K$
and the dephasing rate exhibits an interesting shoulder-peak shape.  
In the narrow boundary region ($\mu \approx \mu_c$), both the resistivity and dephasing rate curves
are similar to the corresponding ones in normal metals. 
The interesting results of the resistivity and dephasing rate are analysized in terms of the calculated spectral
function, self-energy due to Coulomb correlation and also effective impurity level.
The calculated resistivity in the vicinity of $\mu \approx \mu_c$      
is in good agreement with the conventional Kondo like resistivity observed in the recent experiments, 
although the interesting features in the resistivity in the other cases ($\mu < \mu_c$ or $\mu > \mu_c$)
predicted here were not seen in the experiments.                     
We hope that the unusual features of the Kondo resistivity and dephasing rate in graphene reported here
would stimulate further transport experiments on the Kondo effect in graphene.

\section*{Acknowledgments}
The authors thank Ya-Fen Hsu, Chung-Han Wang, and Jong-Chin Lin for fruitful discussions. Supports for this 
work from the National Science Council and the National Center for Theoretical Sciences of Taiwan 
as well as the Deutsche Forschungsgemeinschaft under Grant No. AN 275/7 are gratefully acknowledged.


\begin{references}
\bibitem {Nor} The resistivity for the normal metal Kondo system is calculated by assuming the
scattering rate $\Gamma(\omega)$ is constant [i.e., $\Gamma(\omega) = \Gamma_0$]. 
The impurity level $\epsilon_d = -1.5$ eV used is determined by letting its occupation number ($n$) be 
approximately equal to that in the graphene system with $\mu$ = 0.1 eV [see the inset in Fig. \ref{Resistivity}(a)].
The other parameters are the same as in the graphene case, i.e., $\Gamma_0$ = 1 eV, $U$ = 10 eV and $D = 20$ eV.
\end{references}
\end{document}